\documentclass{aa}
\usepackage{graphicx}
\usepackage{txfonts}
\usepackage[colorlinks]{hyperref}

\hypersetup{
    citecolor=[rgb]{0.04, 0.20, 0.58},  
    linkcolor=[rgb]{0.86, 0.078, 0.235},
    urlcolor=[rgb]{0,0,1}
}
\makeatletter
\renewcommand*\aa@pageof{, page \thepage{} of \pageref*{LastPage}}
\makeatother

\begin{document} 

   \title{Self-supervised component separation for the extragalactic submillimetre sky}

   \author{V. Bonjean\inst{1, 2, 3}
            \and
            H. Tanimura\inst{4, 3}
            \and
            N. Aghanim\inst{3}
            \and
            T. Bonnaire\inst{5}
            \and
            M. Douspis\inst{3}
          }

   \institute{
Littoral, Environnement et Sociétés, Université de La Rochelle, and CNRS (UMR7266), La Rochelle, France
\and
      University of La Laguna, E-38206 Tenerife, Spain.\\
Instituto de Astrof\'{i}sica de Canarias, E-38205 Tenerife, Spain and University of La Laguna, E-38206 Tenerife, Spain.
   \and           
   Université Paris-Saclay, CNRS, Institut d'Astrophysique Spatiale, 91405, Orsay, France \\
   \email{victor.bonjean40@gmail.com}
    \and 
     Kavli Institute for the Physics and Mathematics of the Universe (Kavli IPMU, WPI), University of Tokyo, Chiba 277-8582, Japan
   \and
     Laboratoire de Physique de l’École normale supérieure, ENS, Université PSL, CNRS, Sorbonne Université, Université Paris Cité, F-75005 Paris, France
             }

   \date{Received XXX; accepted XXX}

  \abstract
{
We use a new approach based on self-supervised deep learning networks originally applied to transparency separation in order to simultaneously extract the components of the extragalactic submillimeter sky, namely the cosmic microwave background (CMB), the cosmic infrared background (CIB), and the Sunyaev-Zel'dovich (SZ) effect. In this proof-of-concept paper, we test our approach on the WebSky extragalactic simulation maps in a range of frequencies from 93 to 545 GHz, and compare with one of the state-of-the-art traditional methods, MILCA, for the case of SZ. We first  visually compare  the images, and then  
statistically analyse the full-sky reconstructed high-resolution maps with power spectra. We study the contamination from other components with cross spectra, and particularly emphasise the correlation between the CIB and the SZ effect and compute SZ fluxes around positions of galaxy clusters. The independent networks learn how to reconstruct the different components with less contamination than MILCA. Although this is tested here in an ideal case (without noise, beams, or foregrounds), this method shows significant potential for application in future experiments such as the Simons Observatory (SO) in combination with the \textit{Planck} satellite.
}

   \keywords{methods: data analysis, (cosmology:) large-scale structure of Universe}

   \maketitle


\section{Introduction}

The first light ever propagated in our Universe, namely the cosmic microwave background (CMB), is still detectable today and remains a major probe in observational cosmology. Its spatial anisotropies have been observed and analysed over recent decades by many instruments such as COBE \citep{smoot1992, mather1994}, Boomerang \citep{lange2001}, and WMAP \citep{bennett2003}. The most up-to-date full-sky picture of the CMB was observed in the 2010s by the Planck satellite \citep{planck_hfi2011}, while more detailed CMB features are being observed nowadays in smaller portions of the sky by ground-based experiments such as ACT \citep{fowler2010}, Advanced ACTPol \citep{advact}, SPT \citep{carlstrom2011}, and SPTpol \citep{sptpol}. All these instruments have led to unprecedented precision on the computation of the six cosmological parameters of the concordance $\Lambda$CDM model, which include $\Omega_\mathrm{M}$ and $\sigma_8$ \citep{aghanim2020, aiola2020, aylor2017}. In the coming years, other experiments with even greater spatial resolution, such as the Simons Observatory \citep[SO,][]{ade2019} and CMB-S4 \citep{abazajian2016}, will provide a huge amount of new data and will lead to exciting times for data analysis and cosmology with the CMB.

In the submillimetre frequencies, where the CMB is observable, other sources of emission are also present,  which makes it difficult to disentangle the different components. Some are emissions from objects between our detectors and the CMB; these are the so-called foregrounds, containing for example dust from our galaxy, radio emission, molecular gas emission (CO), and the integrated diffuse infrared emission from all galaxies: the cosmic infrared background \citep[CIB,][]{dole2006}. Other components are distortions of the CMB itself due to its interaction with the objects within the path of the photons; one such distortion is the Sunyaev-Zel’dovich effect \citep[SZ,][]{sunyaev1970}. The known spectral signatures of the different components can potentially be used to separate them, within limitations.

Over recent decades, several algorithms have been developed to separate the different components or clean the CMB; for example, the spectral matching independent component analysis \citep[SMICA,][]{delabrouille2003, cardoso2008} and the generalized morphological component analysis \citep[GMCA,][]{bobin2007}, as well as the sparsity-based algorithms \citep{bobin2008, bobin2013}, Commander \citep{eriksen2008}, needlet internal linear combination \citep[NILC,][]{delabrouille2009}, generalized needlet internal liner combination \citep[GNILC,][]{remazeilles2011}, spectral estimation via expectation maximisation \citep[SEVEM,][]{leach2008, fernandez2012}, modified internal linear combination algorithm \citep[MILCA,][]{hurier2013}, and reduced wavelet scattering transform \citep[RWST,][]{allys2019}. These have been successfully applied in order to separate the CMB \citep[e.g.][]{planck2018i}, but have  also been used to differentiate between components, such as the SZ effect, in different experiments; for example, in \textit{Planck} \citep[][]{planck_sz2016, tanimura2022b}, ACT \citep{madhavacheril2020}, SPT \citep{bleem2022}, and a combination of them such as \textit{Planck}+ACT \citep[PACT,][]{aghanim2019, naess2020} and SPT+\textit{Planck} for CMB lensing \citep{omori2017}. However, for the reconstruction of the SZ effect in particular, one of the main challenges is to confront the contamination from the CIB \citep[e.g.][]{hurier2013, planck_szcib2016}. The CIB signal at high frequencies often leaks in SZ spectral filters and can potentially induce extra power at small scales in SZ power spectra, and vice versa. This is an important effect to take into account when computing cosmological parameters with the SZ power spectrum \citep[e.g.][]{salvati2018,douspis2022,gorce2022,  tanimura2022b}. Another important source of contamination in any CMB component-separation analysis is the galactic dust and the CIB in polarisation data, specifically when aiming to detect the B-mode in polarization \citep{allys2019, lenz2019, regaldo2020}, which is the main future challenge in CMB data analysis.

Deep learning networks, and especially UNets \citep{ronneberger2015}, are very sensitive to both spectral and morphological information and can capture highly non-Gaussian distributions (such as some of the components in CMB frequency maps). This makes those networks particularly suited to studying CMB data and indeed they have been applied with increasing frequency over recent years for different purposes, such as galaxy cluster detection via the SZ signal \citep[e.g.][]{bonjean2020, lin2021}, inpainting of the CMB signal \citep[e.g.][]{puglisi2020, montefalcone2021}, detection of the kinetic SZ (kSZ) effect \citep[e.g.][]{tanimura2022b}, galaxy cluster mass estimations in CMB frequency maps \citep[e.g.][]{gupta2020, deandres2022}, and even foreground cleaning or component separation \citep[e.g.][]{caldeira2019, grumitt2020, petroff2020, lin2021, hurier2021, li2022, wang2022}; also, see \citet{dvorkin2022} for a review. In some cases of machine learning applications, the networks can perform very well (even better than standard methods or than a human) with the condition that very robust, balanced, and well-labelled data is available. Without any known labels ---as in our case, where {the exact different component maps} (SZ, CIB, CMB) are not familiar to us---, self-supervised learning can be used instead. In this kind of machine learning approach, the output Y is reconstructed from the input X without any need for human labelling. Component separation in CMB data is very similar to, for example, transparency separation, dehazing, or mixture images, to which the above-mentioned self-supervised deep learning networks have recently been successfully applied \citep[e.g.][]{gandelsman2018}.

In this paper, we perform component separation of all components of the extragalactic sky simultaneously and in an unsupervised way (specifically, in a self-supervised way). The network, the design of which is inspired by techniques used in transparency separation applied on images \citep{gandelsman2018}, does not rely on known, labelled data for training and could be trained either on numerical simulations to check the performance or directly on real data. Here we apply our algorithm to high-resolution numerical simulations of the extragalactic submillimetre sky from WebSky \citep{stein2020}. Maps are input and reconstructed in HEALPIX format \citep{gorski2005}, with $\mathrm{n}_\mathrm{side}=4096$. Contrary to the methods developed to work in the HEALPIX 1D vector \citep{perraudin2019, krachmalnicoff2019}, we perform the training on small projected patches in two dimensions. We are able to efficiently and simultaneously recover all the input signals (CMB, CIB, and SZ), without any strong contamination from the other components. In this proof-of-concept paper, we demonstrate  the potential of this method by applying it to an ideal case of numerical simulation and by comparing the results ---when possible--- with a state-of-the-art traditional method, MILCA \citep[used to construct the \textit{Planck} $y$ maps;][described in Sect.~\ref{sect:milca}]{hurier2013, planck_sz2016, tanimura2022a}.

Our paper is organised as follows: in Section 2, we present the component maps from the WebSky simulations that we used and describe how we constructed mock maps of the total submillimetre emission at different frequencies, together with MILCA. In Section 3, we present the method, the network architecture, and the training procedure   in detail. We present our results in Section 4, performing different kinds of comparison between the reconstructed component and the original ones, first visually, and then in more quantitative ways, analysing full-sky power spectra, cross-spectra, and SZ fluxes from clusters. In Section 5, we discuss our results and present our conclusions and perspectives for future studies.

\section{Data}\label{sect:data}

In this section, we describe the data used in the present study, that is, the WebSky extragalactic simulation maps from \cite{stein2020} and the derived frequency emission maps.

\subsection{WebSky extragalactic component maps}

The WebSky extragalactic simulation full-sky maps  are a modelisation of the extragalactic components of the submillimetre sky in HEALPIX format, with $n_\mathrm{side}=4096$. They include  maps of the infrared emission (CIB) from dusty star-forming galaxies from $z=0$ to $z=4.6$, a map of the thermal SZ emission from groups and clusters of galaxies, a map of the kinetic SZ effect (kSZ) produced by the Doppler boosting by Thomson scattering of the CMB by bulk flows, and a map of weak gravitational lensing of primary CMB anisotropies by the large-scale distribution of matter in the Universe. The maps are constructed based on a light-cone projection on the full sky of a simulation of halos computed with ellipsoidal collapse dynamics and Lagrangian perturbation theory in the redshift range $0 < z < 4.6$ (with a volume of $\sim600$ ${\left(\mathrm{Gpc}/h\right)}^3$ resolved with $\sim1012$ resolution elements). The distribution of halos is then converted into intensity maps of the different components using models based on existing observations and on hydrodynamical simulations (see \cite{stein2020} for details). The WebSky maps and halo catalogues are publicly available\footnote{\url{https://mocks.cita.utoronto.ca/index.php/WebSky_Extragalactic_CMB_Mocks}}. The WebSky collaboration also provides the conversion factors to apply to the CIB maps and to the SZ map in order to reconstruct the full modelisation of the three components (SZ, CIB, and CMB) on the sky per frequency in units of $\mu K_\mathrm{CMB}$. These latter authors modelled the components in 12 frequencies   in total from SO and \textit{Planck} High Frequency Instrument (HFI), that is, at 27, 39, 93, 100, 143, 145, 217, 225, 280, 353, 545, and 857 GHz. In our study, we discard the frequencies where the radio emission or the dust emission from extragalactic objects are dominant, and so we focus on nine frequencies between 93 GHz and 545 GHz \citep[see e.g. Fig.~4 in][]{planck2018i}. We use the maps of the different components (here SZ, CMB, and CIB) to model the extragalactic submillimetre sky at each frequency in Sect.~\ref{sect:model_maps}, and we use the catalogue of halos to construct a pixel weight map $w$ in Sect.~\ref{sect:halo_map}.

\subsection{Mock submillimetre sky maps}\label{sect:model_maps}

In this study, we focus on extragalactic signals. Therefore, all emissions from our galaxy (i.e. CO, galactic dust, and synchrotron) are not taken into account to model the total emission maps per frequency. We do not take into account either the noise or the effects of beams of the instruments. These effects will be addressed in a future paper. We thus use the components of the WebSky extragalactic simulation maps, namely CMB, SZ, and CIB, at 93, 100, 143, 145, 217, 225, 280, 353, and 545 GHz.
The observed maps $\mathrm{C}_i$ at frequency $i$ can then be expressed in $\mu \mathrm{K}_\mathrm{CMB}$ as
\begin{equation}\label{eq:model}
\mathrm{C}_i = 1\times \mathrm{CMB} + f_i \times \mathrm{SZ} + \mathrm{CIB}_i,
\end{equation}
where $f_i$ are the frequency-dependent weights of the SZ effect and $\mathrm{CIB}_i$ is the CIB in the $i^{\mathrm{th}}$ frequency. As mentioned in Sect.1, all the maps are constructed in HEALPIX format with $n_\mathrm{side}=4096$.

\subsection{Weight map from bright clusters}\label{sect:halo_map}

We focus in particular on the SZ effect and on its correlation with the CIB. As the SZ effect is mainly produced from galaxy clusters, the coverage on the sky is very small; that is, only a very small percentage of the pixels of the HEALPIX maps actually contain a significant SZ signal. As the error on the reconstruction in the case of a UNet approach is based on the mean of all the pixels, we rebalance the statistics of the pixels by constructing a map of weights, $w$, in order to emphasise the regions associated with galaxy clusters. A catalogue of the coordinates on the sky of the halos from the numerical simulations of WebSky was also delivered by \cite{stein2020}, together with some physical properties, such as the redshift $z$ and the mass $M_{200}$. We use this catalogue to compute the SZ flux  for all clusters in the WebSky SZ map and select only the ones with an SZ flux detected with more than $5\sigma$ from a distribution of fluxes computed at random positions on the sky. We select only the brightest ones in order to apply a strategy that could also be applied in real data (in which we would have only the brightest SZ clusters with which to apply weights). We construct the weight map $w$ in HEALPIX, with 1 at the position of the approximately $13,000$ selected clusters, and put 0 otherwise. We then convolved the map with an FWHM of 5 arcmin (chosen arbitrarily; we further checked that changing this value does not affect the results), and squared the weight-map to further emphasise the central pixels of clusters. We divided by the sum of the pixels so that the average of the map on the pixels is equal to 1, as in the case of a uniform weight map. By weighting the pixels is this way, the very central regions of the clusters account for about 50\%, and the remaining 50\% are associated with the external regions. We later show that this weighting procedure simply helps the network to converge faster but is not dependent on the catalogue of clusters for which we weight the pixels. As all the clusters in the real sky are not known, this is a very important statement, and subsequently allows us to apply this self-supervised method to real data.

\subsection{MILCA maps}\label{sect:milca}

We use MILCA-based reconstructed maps of SZ and CMB applied in the WebSky submillimetre maps as a reference comparison in our study. MILCA \citep{hurier2013,tanimura2022a} is based on the internal linear combination (ILC) approach, which preserves an astrophysical component given the known spectrum by minimising the variance in the reconstructed signal. In MILCA, extra degrees of freedom are used to null-out other components (such as CMB for SZ) and minimize noises using noise maps estimated from split maps such as half-mission maps. This approach was used to extract the SZ signal using \textit{Planck} data \citep{planck_sz2016, tanimura2022a} and \textit{Planck}+ACT data \citep{aghanim2019}. However, the correlation between SZ and CIB is known to induce a certain level of contamination in the resulting $y$-map \citep{planck_szcib2016}, especially at small scales, and causes a large uncertainty in astrophysical \citep{vikram2017} and cosmological analyses \citep{hill2014, planck_sz2016, tanimura2022a}.

\section{Method}\label{sect:methods}

In this section, we explain our model of the submillimetre extragalactic sky, the architecture of the network used, and the construction of the training set we use.

\subsection{Model}

The aim of the study is to separate the CMB, the SZ, and the CIB emissions  as efficiently as possible using both frequency and spatial morphologies. Unlike the CMB or the SZ emissions, the weights of the CIB spectrum are different at each line of sight, meaning that there should be as many CIB maps as frequencies used. To perfectly recover all the components, one should then recover (i) the CMB, (ii) the SZ $y$ map, and (iii) the CIB at all the frequencies used, which gives $2+n$ components  (where $n$ is the number of frequency used, here $n=9$);  meaning a total of 11 in our case. However, it is impossible to recover more components than the number of maps in inputs because of the important degeneracies between solutions. Hence, we assume that the CIB in the different frequencies, $\mathrm{CIB}_i$, can be approximated by one CIB map fixed at one frequency $\overline{\mathrm{CIB}}$ (here the maximum frequency $f= 545$ GHz) weighted by the values $\psi_i(z),$ which represent a modified black body per pixel at a mean redshift $\overline{z}$. This enables the frequency dependency of the weights, and both the maps $\overline{\mathrm{CIB}}$ and $\overline{z}$ to be learned at the same time during the training. The model can be written as
\begin{equation} \label{eq:cib}
    \mathrm{CIB}_i = \psi_i\left(\overline{z}\right) \times \mathrm{\overline{CIB}},
\end{equation}
with
\begin{equation}
    \psi_i\left(\overline{z}\right) = \left(\frac{i}{545}\right)^{\beta + 3} \times \frac{\mathrm{exp}\left(\frac{h\times 545\times10^9 (1+\overline{z})}{k_\mathrm{B}\mathrm{T}_0 (1+\overline{z})^\alpha}\right)-1}{\mathrm{exp}\left(\frac{h\times i\times 10^9 (1+\overline{z})}{k_\mathrm{B}\mathrm{T}_0 (1+\overline{z})^\alpha}\right)-1},
\end{equation}
where $\beta$ is the slope of the power law of the modified black body, $\mathrm{T}_0$ is the effective dust temperature, $\alpha$ is the parameter for the redshift dependency of the dust temperature, $h$ is the Planck constant, and $k_\mathrm{B}$ is the Boltzmann constant. In this study, we fixed $\beta=1.6$, $\mathrm{T}_0 = 20.7$ K, and $\alpha=0.2$ following \cite{stein2020}. While it is a rough approximation to consider the CIB as a modified black body with a fixed $\beta$ per pixel (i.e. per line of sight), it is still a reasonable approximation for this work. In a forthcoming analysis, we will consider more realistic models of the spectrum of the CIB using for example the Taylor expansion of the modified black body proposed in \cite{chluba2017}.

Merging Eq.~\ref{eq:model} and Eq.~\ref{eq:cib}, the maps $\mathrm{C}_i$ can then be modelled by
\begin{equation} \label{eq:model2}
    \overline{\mathrm{C}_i} = 1\times \overline{\mathrm{CMB}} + f_i \times \overline{\mathrm{SZ}} + \psi_i\left(\overline{z}\right) \overline{\mathrm{CIB}},
\end{equation}
where $\overline{\mathrm{CMB}}$ is the extracted CMB component, $\overline{\mathrm{SZ}}$ is the reconstructed SZ component, $\overline{\mathrm{CIB}}$ is the extracted CIB component at 545 GHz, and $\overline{z}$ is the spatially dependent map of the mean redshift of the modified black body of the CIB per pixel. Here, we simultaneously fit the four components $\overline{\mathrm{CMB}}$, $\overline{\mathrm{SZ}}$, $\overline{\mathrm{CIB}}$, and $\overline{z}$, still allowing nine different CIBs.

\subsection{Network architecture}\label{sect:archi}

\begin{figure}[!ht]
\centering
\includegraphics[width=0.5\textwidth]{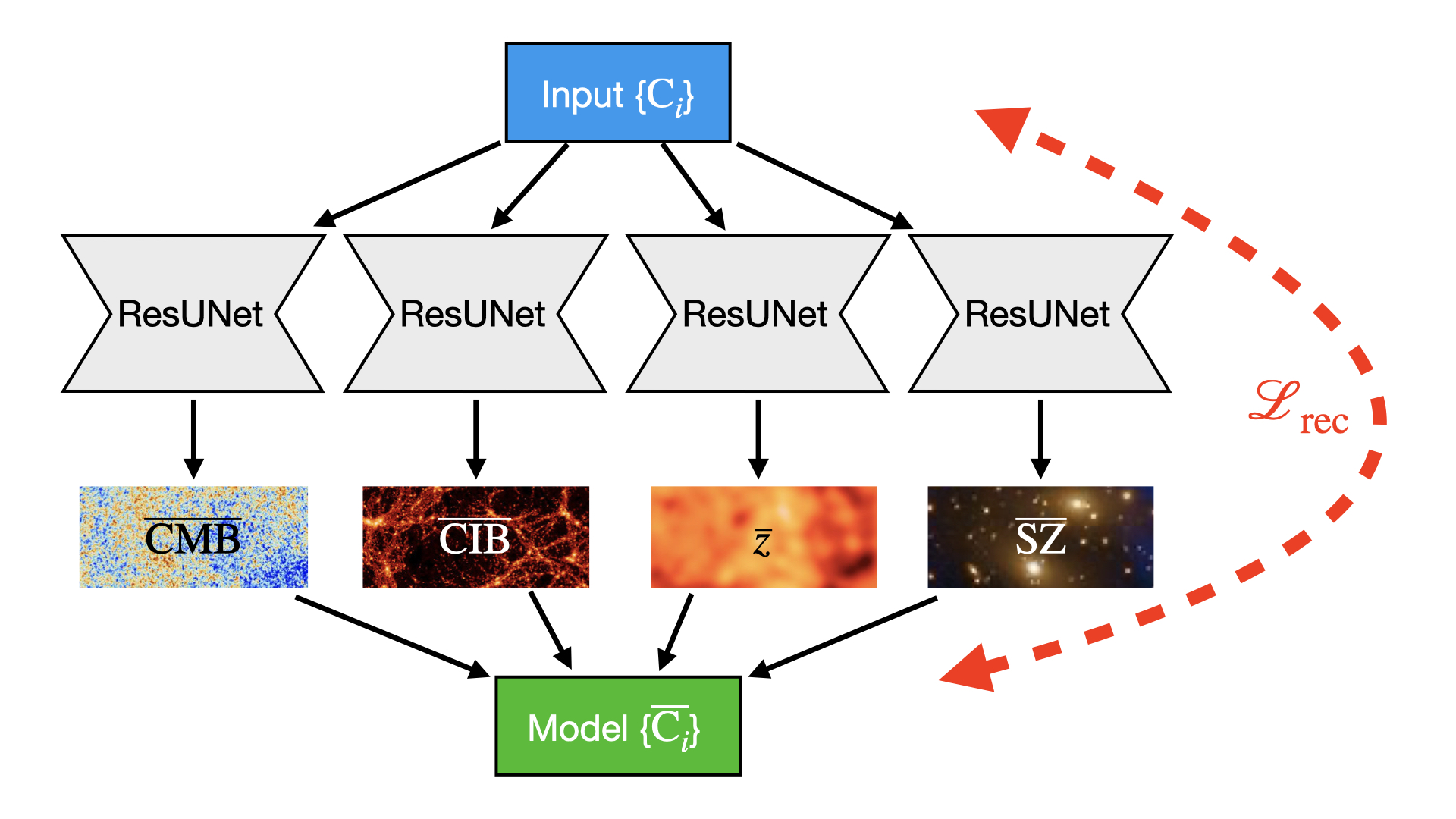}
\caption{\label{fig:schema}Diagram of our architecture. Four ResUNets are used to construct $\overline{\mathrm{CMB}}$, $\overline{\mathrm{SZ}}$, $\overline{\mathrm{CIB}}$, and $\overline{z}$ directly from the input $\mathrm{C_i}$, and a reconstruction loss $\mathcal{L}_\mathrm{rec}$ is applied between the inputs $\mathrm{C_i}$ and the models $\overline{\mathrm{C_i}}$ obtained with Eq.~\ref{eq:model2}.}
\end{figure}

In this study, we use a combination of specific architectures of deep convolutional neural networks (CNNs), the so-called ResUNets ---which have been shown to be very efficient in image regression \citep[][]{zhang2018}---, to reconstruct the three components $\overline{\mathrm{CMB}}$, $\overline{\mathrm{SZ}}$, and $\overline{\mathrm{CIB}}$ directly from the input maps $\mathrm{C}_i$. Our approach is based on transparency separation for dehazing of images developed in \cite{gandelsman2018}. Indeed, in this domain, they have a very similar goal; they are all designed to reconstruct a blurred image by decomposing it into a dehazed image and an image of the haze. Another application is the recovery of two images that are superposed on each other, or the removal of water droplet marks on the camera lens and seen in pictures. This technique is quite simple: the goal is to output two images from one, reconstructed by two different networks in parallel (allowing the non-correlation between the two images as each network is independent from the other), and by adding a term in the loss function that minimises the cross-correlation between the two reconstructed images. This technique has shown surprisingly good results, especially in the domain of dehazing images \citep{gandelsman2018, feng2021, wenyi2021, miao2022}. This is somewhat similar to separation techniques for emission in submillimetre. We apply our knowledge of the physics concerning the frequency dependency of the components  to the outputs of the ResUNets in
parallel (that are images) in order to construct a final combined output $\overline{\mathrm{C}_i}$ (given by Eq.~\ref{eq:model2}) that has to match the input maps. We apply a loss function $\mathcal{L}_\mathrm{rec}$ between the final outputs $\overline{\mathrm{C}_i}$ and the inputs $\mathrm{C}_i$, and the component is extracted by the outputs of each ResUNet. We can put different priors by modifying the activation function on the different outputs of each network. Doing this helps the networks to converge faster to the accurate solution and helps avoid degeneracies. For $\overline{\mathrm{CMB}}$, we choose an activation function that is a Sigmoid between -1000 and +1000 $\mu K_\mathrm{CMB}$. For $\overline{\mathrm{CIB}}$, we choose a Softmax function that outputs only positive numbers (CIB being a positive luminosity). For the redshift of CIB $\overline{z}$, we choose a Sigmoid function between 0 and 5. For $\overline{\mathrm{SZ}}$, we choose a Softmax function removing $-20\times 10^6$ so that $y>-20\times 10^6$. Although we know the $y$ Compton parameter cannot be negative, forcing it to be positive only leads to a non-Gaussian error on the reconstruction map. This effect produces a bias in all the mean values of the different components in an attempt to reproduce Gaussian noise in the total reconstruction of the frequency maps (characteristics of the MSE loss function). A diagram of our architecture is shown in Fig.
~\ref{fig:schema}.

We define the reconstruction loss $\mathcal{L}_\mathrm{rec}$ as the sum of the weighted mean squared errors (wMSE) in the different frequencies $i$ divided by the standard deviation of the maps ${\sigma_i}$, namely
\begin{equation}
    \mathcal{L}_\mathrm{rec} = \frac{1}{n_\mathrm{f}}\sum_{i = 0}^{n_\mathrm{f}} \frac{\mathrm{wMSE}\left(\mathrm{C}_i, \overline{\mathrm{C}_i}\right)}{\sigma_i},
\end{equation}
where $n_\mathrm{f}$ is the number of frequencies (here, $n_\mathrm{f}=9$), $\mathrm{wMSE}\left(\mathrm{C}_i, \overline{\mathrm{C}_i}\right)$ is the MSE in the $i^\mathrm{th}$ frequency weighted by the weight map $w$, and $\sigma_i$ is the standard deviation of the map $\mathrm{C}_i$.

\subsection{Training set}\label{sect:training_set}

Our network architecture is applied to sets of 2D patches projected on the sky. We extracted $n=100,000$ multi-channel 2D patches  from the $\mathrm{C}_i$ maps, with the nine frequencies from 93 to 545 GHz. The images are $64\times64$ pixels with a resolution of $\theta_{\mathrm{pix}}\sim 0.8$ arcmin (giving a field of view of $\sim0.8^\circ \times \sim0.8^\circ$). 

The dimensions of the input data are thus $100,000\times64\times64\times9$ pixels. As in \cite{bonjean2020}, we train our network on 90\% of the sky and leave 10\% of the sky completely unseen by the network (as commonly done in machine learning applications). We also tried a ratio of 80\%-20\% and checked that this choice was not affecting the results. In this first application, the 10\% unseen fraction of the sky is randomly selected on the sphere. One could however use a clean area of the sky (away from galaxy contamination) for a study including dust and radio emission. In this configuration, we can compute global properties of the different extracted components at each epoch of the training, such as the means and the variances, and compare them to the expected values. When those values converge into a good solution and reach a plateau together with the loss value on this very same 10\% unseen area, we consider that the network has reached convergence and stop it. With the trained models, we subsequently reconstruct the full-sky HEALPIX maps   entirely by estimating projected patches on the full sky and averaging the pixels, at $\mathrm{n}_\mathrm{side}=4096$, for the different components.

\section{Results}\label{sect:results}

In this section, we present the results and compare the reconstructed HEALPIX maps of the different extracted components to the original one from WebSky in several ways.

First, we visually inspect the maps by showing projections of the components centred on the brightest SZ cluster to quickly check the contrasts and the non-Gaussian aspects of the pixel distributions of the different fields.
We then present further quantitative statistical comparisons and show the power spectra and the PDF of the pixels of the full-sky maps for
all the components. We also compare the cross power spectra with the original components. For the case of the SZ effect, we also compare with the outputs of a traditional but state-of-the-art method, MILCA \citep[][]{hurier2013, planck_sz2016, tanimura2022a}.
We compute the cross spectra between all the different components to study the contamination of the maps coming from the leakage from other components.
Finally, we put particular emphasis on the regions around galaxy clusters, and perform an SZ flux comparison between those extracted from the reconstructed components and those from the WebSky maps.

\begin{figure*}[!ht]
\centering
\includegraphics[width=0.5\textwidth]{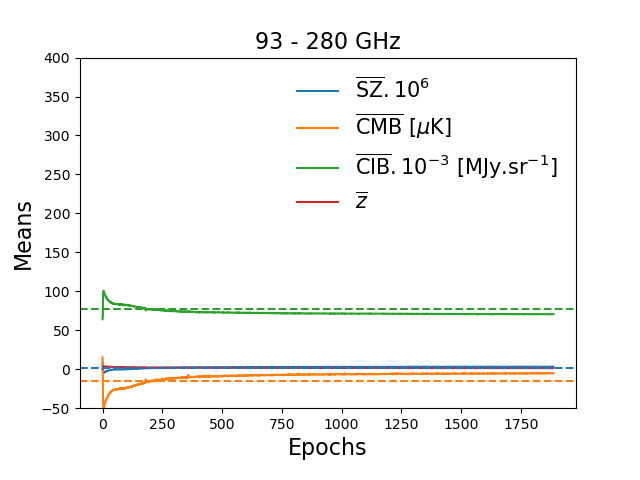}\includegraphics[width=0.5\textwidth]{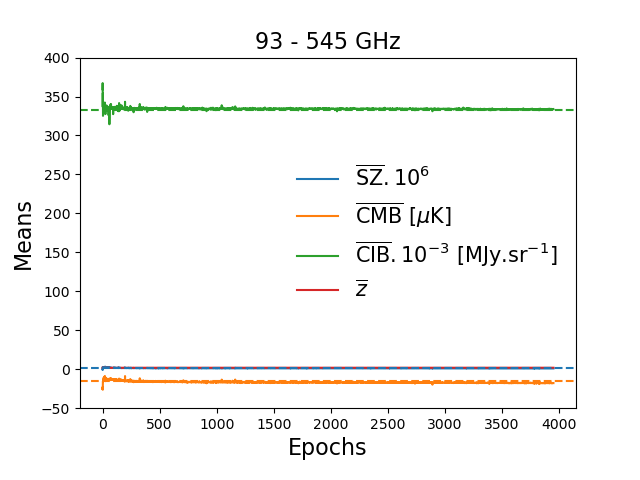}
\includegraphics[width=0.5\textwidth]{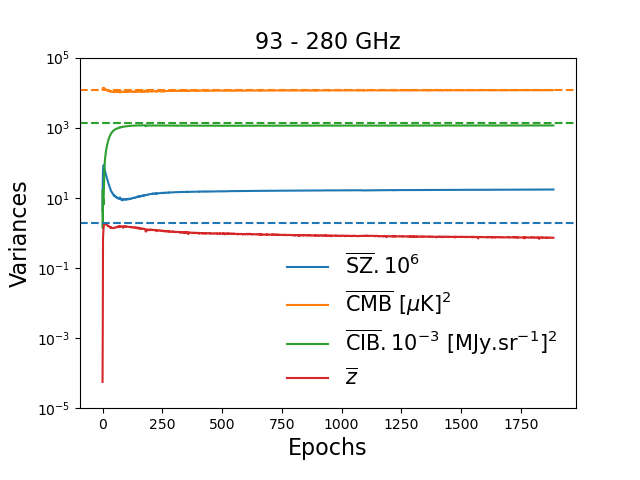}\includegraphics[width=0.5\textwidth]{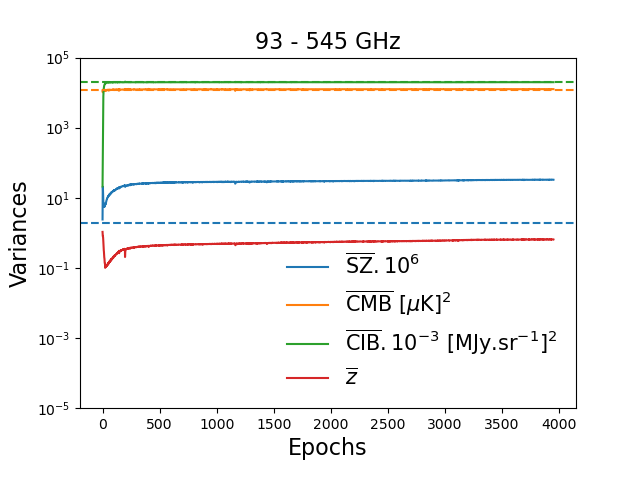}
\caption{\label{fig:ev_prop}Evolution of the different statistical quantities of the extracted components in the unseen area of the sky. Top: Evolution of the means of the components on the left for the seven-frequency configuration and on the right for the nine-frequency configuration. Bottom: Evolution of the variances of the components on the left for the seven-frequency configuration and on the right for the nine-frequency configuration. Dashed lines represent the expected values. The values for the CIB are not the same for the seven and the nine frequencies models, as we are computing the CIB at 280 GHz for the first case and the CIB at 545 GHz for the second case. Units are different for each map: $\overline{\mathrm{SZ}}$ and $\overline{z}$ are without units ($y$ and redshift), $\overline{\mathrm{CMB}}$ is in $\mu \mathrm{K}$, and $\overline{\mathrm{CIB}}$ is in $\mathrm{MJy.sr}^{-1}$.}
\end{figure*}

\subsection{Frequency settings}

First, we explore the impact of the choice of the frequency setting on the results and for this we define two combinations of frequencies: one with seven frequencies, from 93 to 280 GHz, and another combination adding the 353 and 545 GHz frequency maps in which the CIB is dominant. We compared the global properties of the components in the unseen area for each of the epochs and for the different combinations. The means and variances of the reconstructed components for the two frequency combinations are shown in Fig.~\ref{fig:ev_prop}. We directly see that the configuration with seven frequencies is not evolving into an expected solution (shown by the dashed lines), but rather shows biases in the mean values of all components and a lack of variance for the CIB. In the other configuration, the mean and variances are well recovered, except for the variance of SZ, which is expected as SZ is a very weak signal and hence some reconstruction noise is expected,  adding extra variance. This is due to the spectral dependencies of the SZ and the CIB in the range 100 - 250 GHz, which are too similar, making it difficult to break the degeneracy between the two components. Instead of learning a lower redshift $\overline{z}$ of the CIB at the position of the clusters as we expect (as CIB at the positions of clusters is dominated by the galaxies from the galaxy clusters itself), it does the opposite and tends to learn a higher redshift. This produces a lower bias in the SZ flux to balance the total emission and the model hence converges to an inaccurate solution. For these reasons, considering our model, we must take into account the higher frequency maps at 353 and 545 where the CIB is dominant and where the spectral dependencies of the two components SZ and CIB start to differ one from another. In the following, our results are thus obtained in the configuration with nine frequencies, including the 353 and 545 GHz maps. We also note that in Fig.~\ref{fig:ev_prop}, the values of the means and variances of the CIB are not the same for seven and nine frequencies, as we are computing the CIB at 280 GHz for the first case and the CIB at 545 GHz for the second case.

\begin{figure*}[!ht]
\centering
\includegraphics[width=0.33\textwidth]{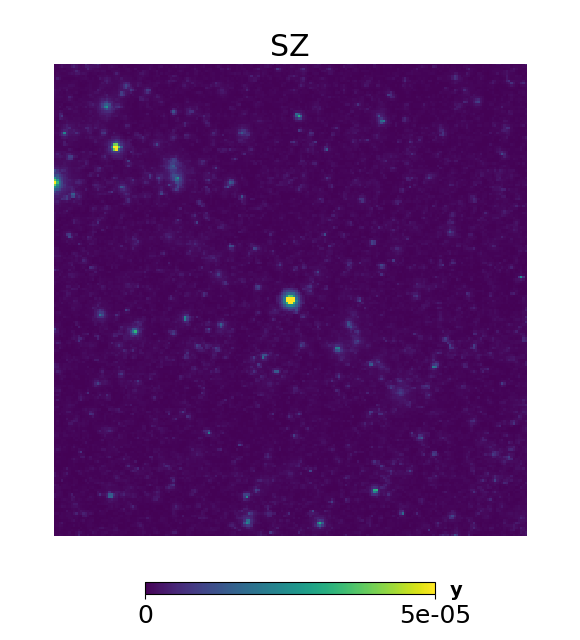}\includegraphics[width=0.33\textwidth]{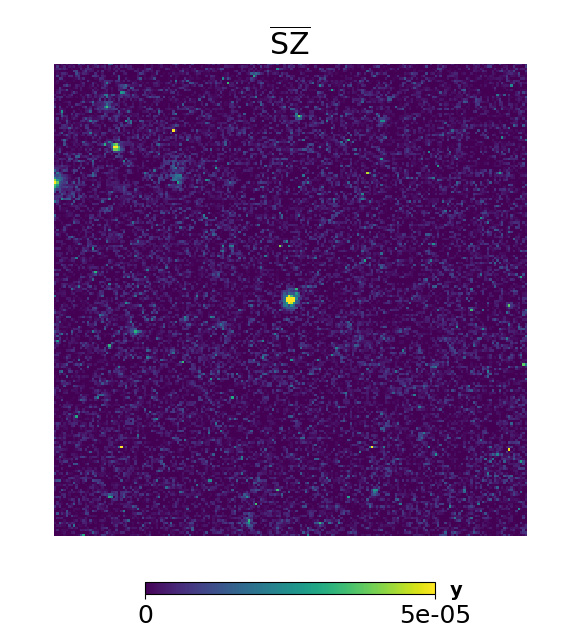}\includegraphics[width=0.33\textwidth]{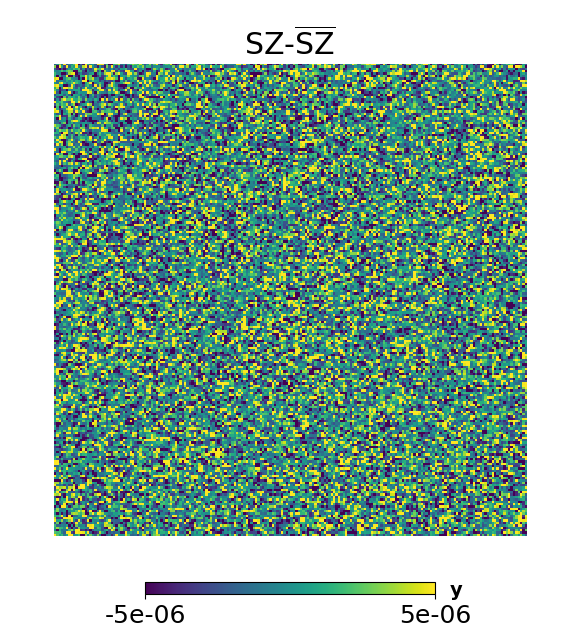}
\includegraphics[width=0.33\textwidth]{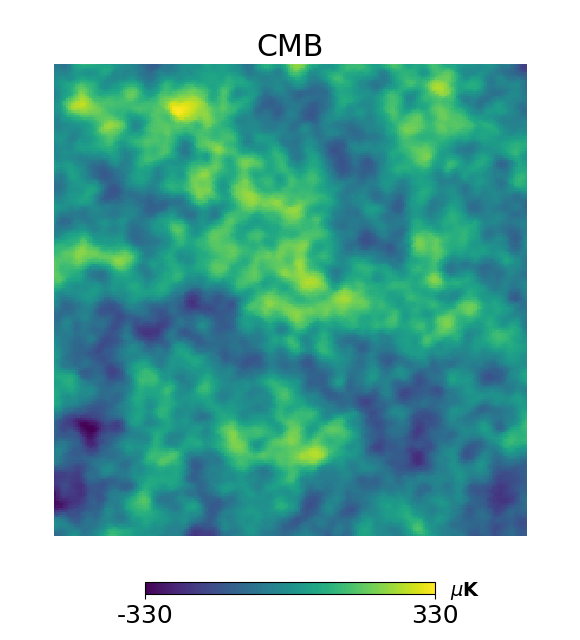}\includegraphics[width=0.33\textwidth]{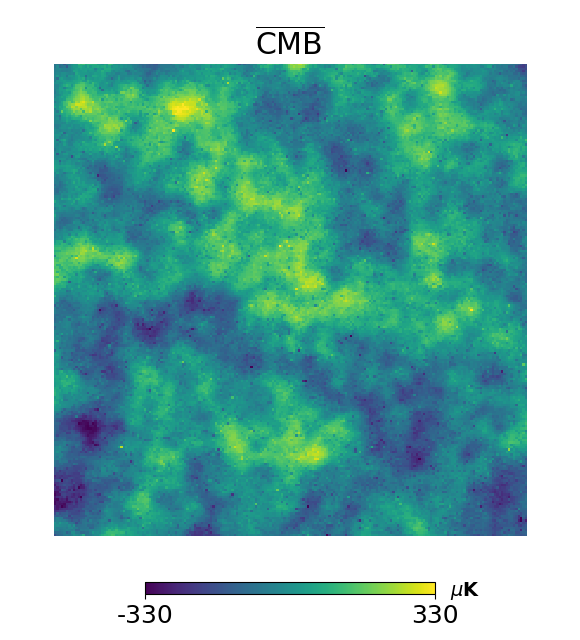}\includegraphics[width=0.33\textwidth]{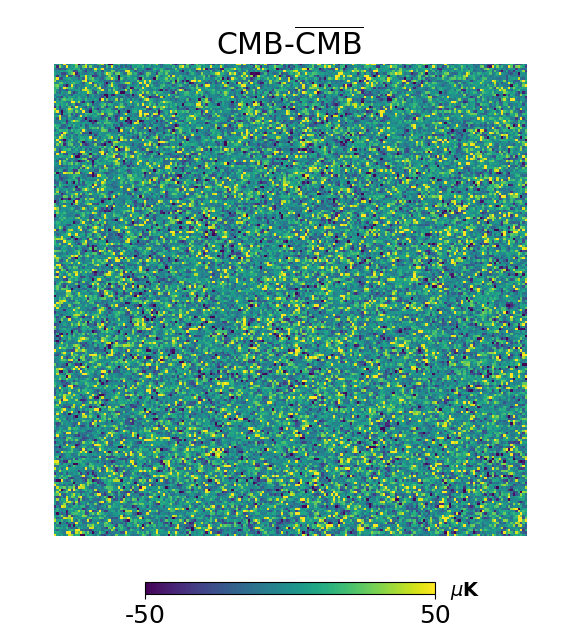}
\includegraphics[width=0.33\textwidth]{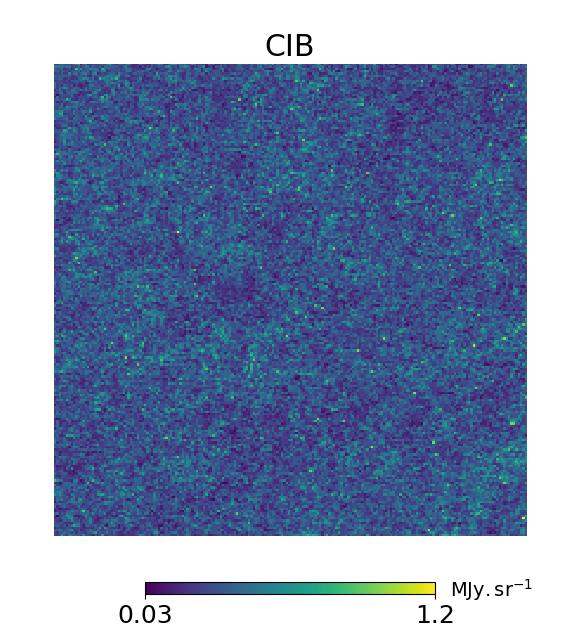}\includegraphics[width=0.33\textwidth]{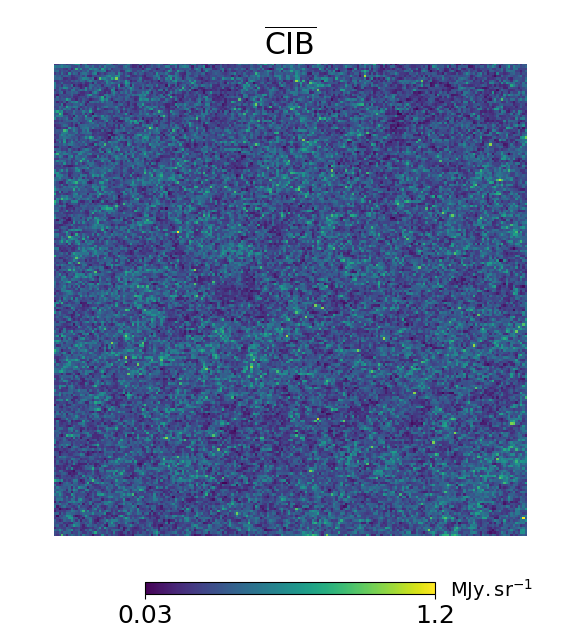}\includegraphics[width=0.33\textwidth]{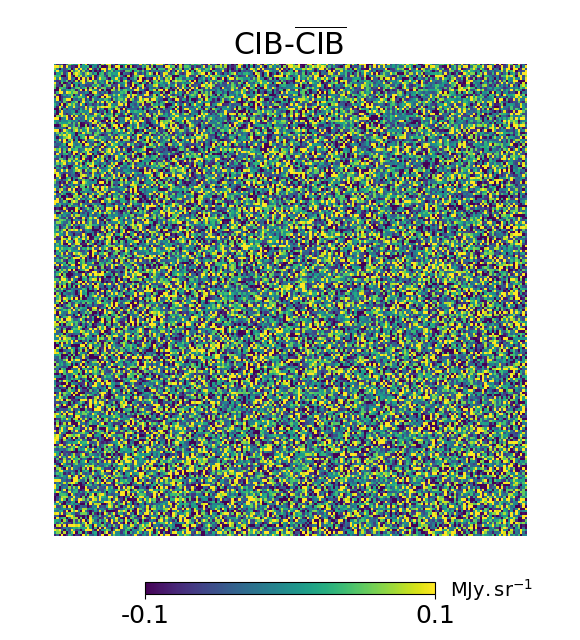}
\caption{\label{fig:proj}Projection in a $5^\circ\times5^\circ$ patch of the HEALPIX maps of the different components around the brightest SZ cluster. Left column: From top to bottom, the SZ, CMB, and CIB from WebSky simulation maps. Middle column: From top to bottom, the reconstructed $\overline{\mathrm{SZ}}$, $\overline{\mathrm{CMB}}$, and $\overline{\mathrm{CIB}}$. Right column: Residuals between the WebSky component and the reconstructed ones. Very good agreement is seen visually between the reconstructed maps and the expected map, without structures in the residuals.}
\end{figure*}

\begin{figure*}[!ht]
\centering
\includegraphics[width=0.33\textwidth]{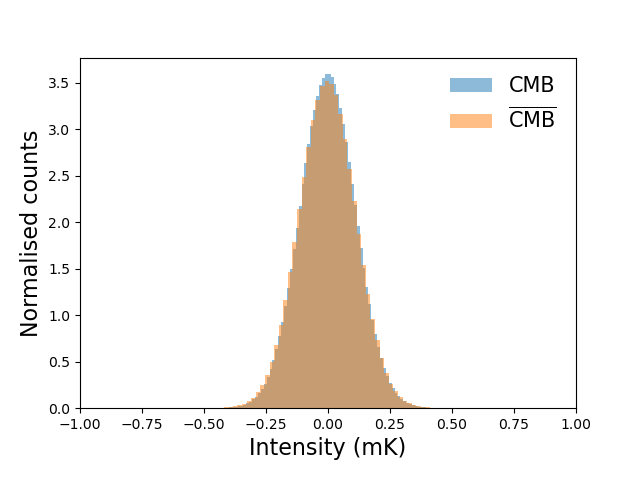}\includegraphics[width=0.33\textwidth]{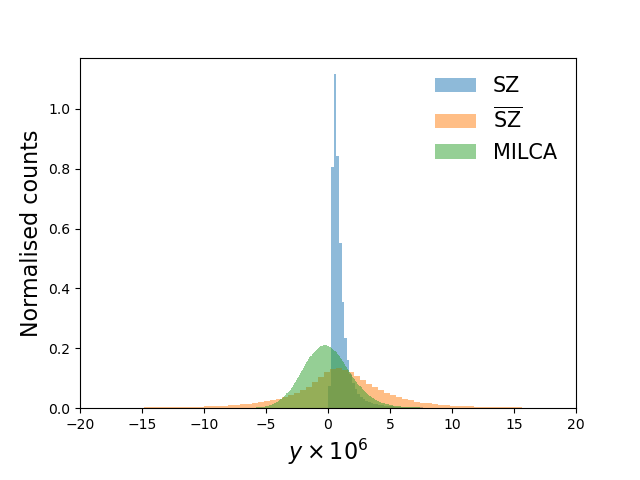}\includegraphics[width=0.33\textwidth]{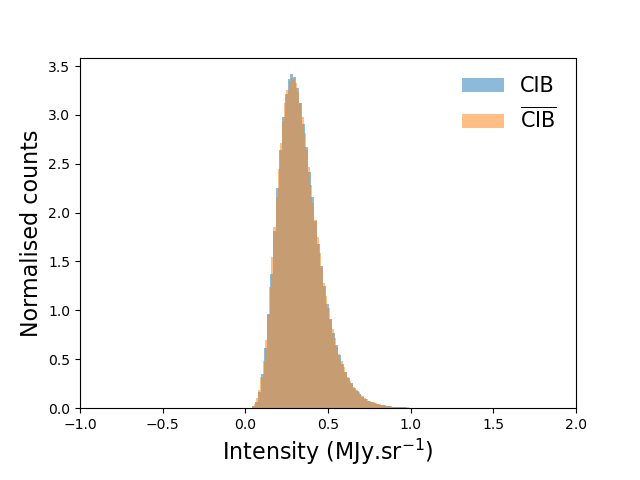}
\caption{\label{fig:pdf}PDF comparison between the reconstructed components and the WebSky ones. Left: Case for the CMB. Middle: Case for the SZ. Right: Case for the CIB.}
\end{figure*}

\begin{figure*}[!ht]
\centering
\includegraphics[width=0.33\textwidth]{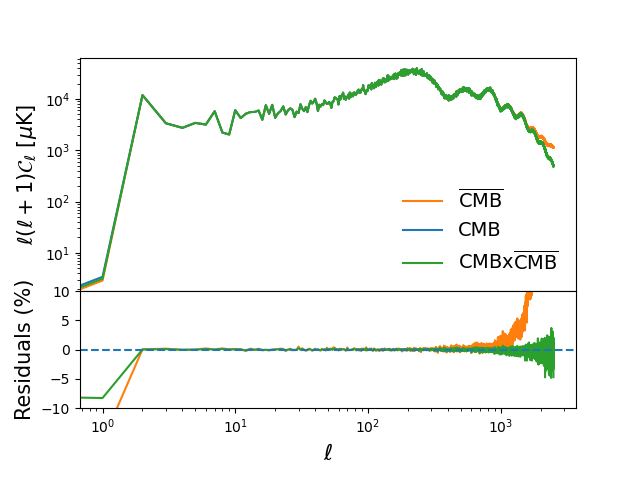}\includegraphics[width=0.33\textwidth]{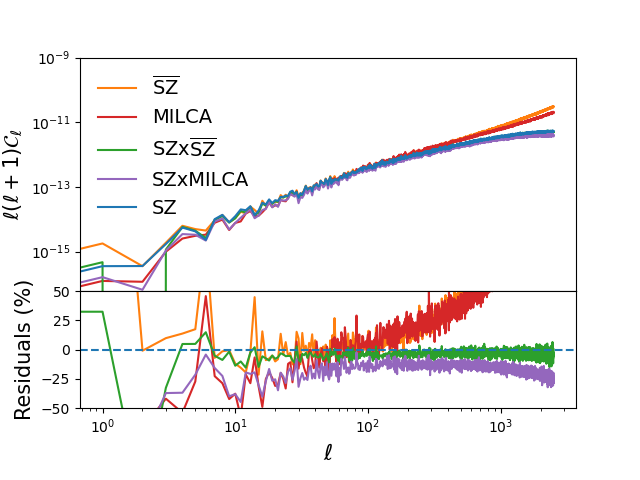}\includegraphics[width=0.33\textwidth]{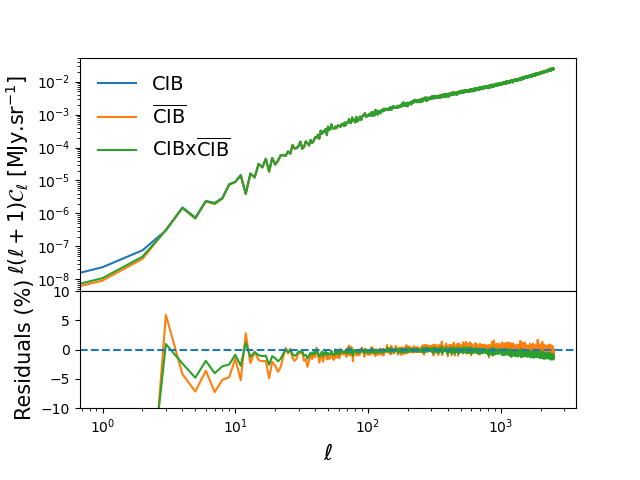}
\caption{\label{fig:powerspectra}Power spectra of the components, the recovered components, and the cross power spectra between the two. $\mathcal{C}_\ell$ residuals as a percentage are also shown in the bottom of each plot, where residuals are defined as $\frac{x-\overline{x}}{x}\times100$ for the quantity $x$. Left: Case for the CMB. Middle: Case for the SZ. MILCA is also shown in this plot. Right: Case for the CIB.}
\end{figure*}

\begin{figure*}[!ht]
\centering
\includegraphics[width=0.33\textwidth]{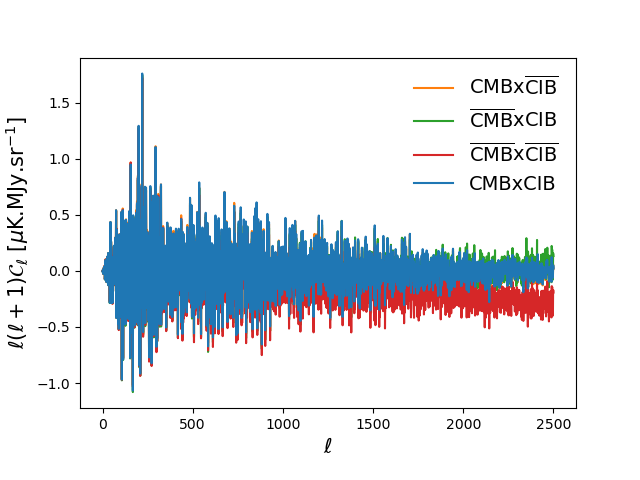}\includegraphics[width=0.33\textwidth]{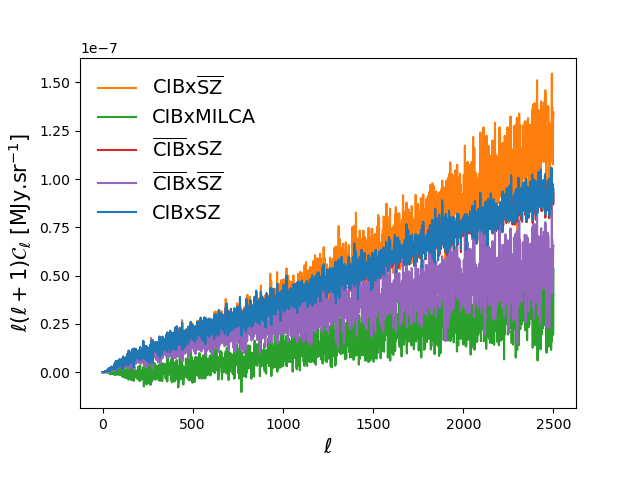}\includegraphics[width=0.33\textwidth]{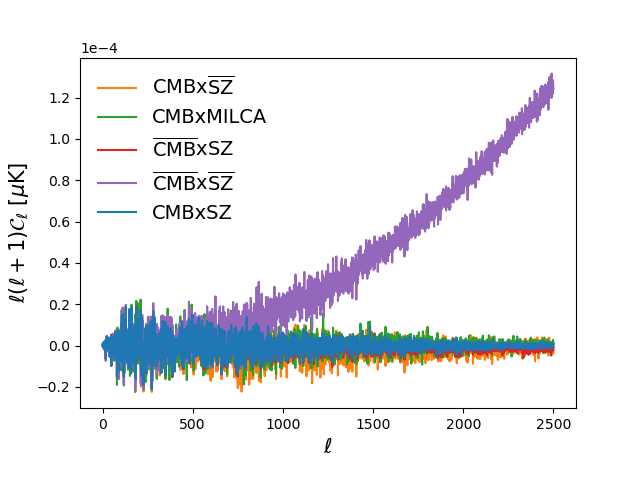}
\caption{\label{fig:crossspectra}Cross power spectra between the different components. Left: Case for CMBxCIB. Middle: Case for CIBxSZ. MILCA is also considered in this plot. Right: Case for CMBxSZ.}
\end{figure*}

\subsection{Visual inspection}

We trained our network with nine frequencies and reconstructed the HEALPIX maps of the four extracted components: $\overline{\mathrm{CMB}}$, $\overline{\mathrm{SZ}}$, $\overline{\mathrm{CIB}}$, and $\overline{z}$. The training lasted six days on a Tesla V100 GPU. A first comparison can be made visually to qualitatively check the ranges of pixel values, contrasts, and the non-Gaussian distribution of the pixels of the fields. We present an example in Fig.~\ref{fig:proj}, where we show the projection of the different maps on a $5^\circ\times5^\circ$ patch centred around the position of the brightest SZ cluster. The left column shows the WebSky maps of CMB, SZ, and CIB; the middle column shows the reconstructed maps $\overline{\mathrm{CMB}}$, $\overline{\mathrm{SZ}}$, and $\overline{\mathrm{CIB}}$, and the right column shows the residuals between the true and reconstructed components. The reconstructed components appear, visually, to be in good agreement with the WebSky original components, with clusters well recovered, and without any significant features seen in the residual images.

\subsection{PDF comparison}\label{sect:pdf}

We pursue our comparison by checking the PDF of the pixels for the different components. In Fig.\ref{fig:pdf}, we show the distribution of the pixels for the $\overline{\mathrm{CMB}}$, $\overline{\mathrm{SZ}}$, and $\overline{\mathrm{CIB}}$, where we compare to the original distributions from WebSky maps. For the case of SZ, we also compare to the distribution of pixels from the SZ maps reconstructed with MILCA. We see a very good match between the distributions of pixels for CMB and CIB, while the distribution is more flattened for the recovered SZ maps from our method and from MILCA. This indicates a noisy reconstruction of the SZ signal, which is expected considering the very low amplitude of the SZ effect compared to CMB or CIB, which are the dominant signals in some frequencies. However, the variance of the reconstructed SZ map with our method seems larger than that of the map obtained with MILCA, which is also expected as our model contains a greater number of free parameters than MILCA (simultaneously constraining CMB, SZ, z, and CIB). This is discussed in more detail when comparing cross-spectra and contamination from other components below.

\subsection{Power-spectra comparison}\label{sect:spectra}

We computed the auto power spectra for all the recovered components and compared them to the power spectra of the original WebSky components. For each component, we also compare the cross power spectra between the recovered components and the original ones to check whether or not  the signal is fully recovered when removing the reconstructed noise of the estimated components. In Fig.~\ref{fig:powerspectra}, we show the resulting power spectra for the CMB in the left panel, SZ in the middle one, and CIB on the right. We also show the $\mathcal{C}_\ell$ residuals as a percentage at the bottom of each plot. For the case of CMB, we recover the signal very well, with reconstructed noise dominating from $\ell>1000$. Removing the reconstructed noise in the cross spectrum with the WebSky CMB leads to very good recovery of the CMB signal, with below 1\% error up to $\ell=2500$. For the CIB, we see very good agreement between the two power spectra, with less than 3\% error and a bias of below 2\% up to $\ell=2500$. The cross spectrum between $\overline{\mathrm{CIB}}$ and CIB also shows a good recovery of the signal. For the SZ reconstructed maps, the power spectra are above the expected signal for both MILCA and our method $\overline{\mathrm{SZ}}$, especially at small scales. This indicates the presence of additional signal ---for example, noise---, which is in good agreement with the effect seen in the distributions of the PDF in Sect.~\ref{sect:pdf} (greater noise for $\overline{\mathrm{SZ}}$ than for MILCA). By performing the cross spectra with the SZ WebSky map, interesting observations can be made. At all $\ell$, we see a lack of signal in the cross spectrum between MILCA and the true SZ emission from WebSky (MILCAxSZ), leading to a bias of between 10\% and 25\% within $10<\ell<2500$, which is not seen in the cross-spectrum $\overline{\mathrm{SZ}}$xSZ. This effect is due to the contamination of the CIB in the MILCA SZ map, itself attributable to the correlation between SZ and CIB. The residuals of the cross-spectrum $\overline{\mathrm{SZ}}$xSZ indicate a good recovery of the signal within an error of 5\% up to $\ell=2500$ and with a bias of below 3\% at all $\ell$. We confirm this result in the following section, where we demonstrate how we compute the cross-spectra between all the components to study the contamination.

\begin{figure*}[!ht]
\centering
\includegraphics[width=0.5\textwidth]{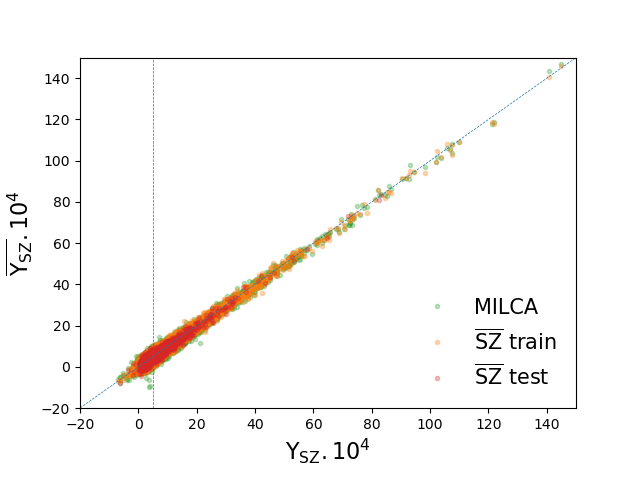}\includegraphics[width=0.5\textwidth]{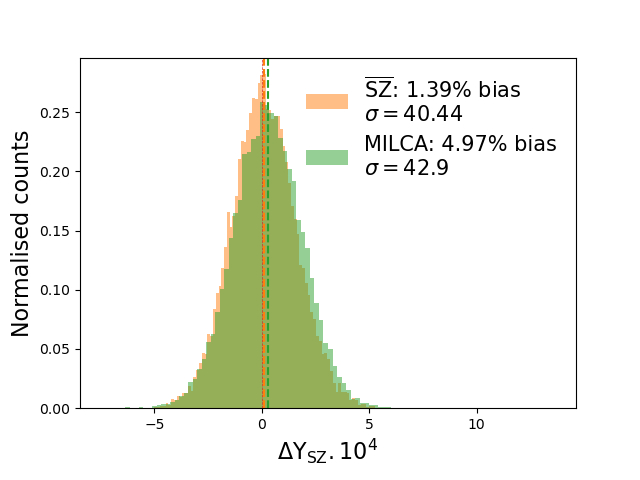}
\caption{\label{fig:fluxes}Comparison of the SZ fluxes computed with aperture photometry in the exact same way in the SZ WebSky map, in the $\overline{\mathrm{SZ}}$ map, and in the MILCA SZ map. Left: Fluxes on the reconstructed maps as a function of the SZ WebSky fluxes. The blue dashed line represents the SZ flux have been enhanced in the weight map $w$. Right: Histograms of the residuals. The dashed lines show the median values of the distributions.}
\end{figure*}

\subsection{Cross-spectra between components}

We computed the cross-power spectra between all the components CMBxSZ, CMBxCIB, and CIBxSZ. In each case, we cross correlated all combinations between the original WebSky components and the reconstructed ones, either from MILCA or with our method for the cases where SZ is involved. Figure~\ref{fig:crossspectra} shows the results for all the different components. We detail the different cases  hereafter.

\subsubsection{CMBxCIB}
For the CMBxCIB case (left panel of Fig.~\ref{fig:crossspectra}), where there is no correlation expected (blue line), we did not find any correlation between the different combinations, except for $\overline{\mathrm{CMB}}$x$\overline{\mathrm{CIB}}$. This means that the reconstructed CMB is not contaminated by the CIB, and that the reconstructed CIB is not contaminated by the CMB. On the other hand, the correlation between $\overline{\mathrm{CMB}}$ and $\overline{\mathrm{CIB}}$ is expected (but not necessarily problematic), as this correlation is coming from the correlated reconstructed noise between the two components, which is not expected to be independent, as the two components are reconstructed at the same time. 

\subsubsection{CIBxSZ}
For the CIBxSZ  case (middle panel of Fig.~\ref{fig:crossspectra}), there is an expected correlation between the two components (shown in blue) \citep[][]{planck_szcib2016, stein2020}. This correlation comes from the fact that the dominant part of the CIB in clusters, which is where the SZ signal is dominant, is generated from the dust emission of galaxies in these very same clusters. This correlation is a bias in the SZ reconstruction \citep[e.g.,][]{planck_szcib2016} that we recover in $\overline{\mathrm{CIB}}$xSZ (in red), indicating that the reconstructed CIB is not contaminated by the SZ. However, CIBxMILCA (in green) is below the correlation line, indicating that the MILCA SZ map is perturbed by the CIB and might lack flux. This translates into a lack of SZ power at small scales, which is also seen in the SZ auto power spectra in Sect.\ref{sect:spectra}. However, the cross-spectrum CIBx$\overline{\mathrm{SZ}}$ (in orange) is closer to the expected correlation in blue, indicating a weaker contamination from the CIB in the $\overline{\mathrm{SZ}}$ reconstructed map. In conclusion, the reconstructed $\overline{\mathrm{SZ}}$ map is noisier than that produced with MILCA (with noise coming from the reconstruction method and larger than that produced by MILCA, as our model contains a greater number of free parameters, as it simultaneously constrains CIB, SZ, z, and CIB; this would also be the case for MILCA (noise would increase if we increase the number of constraints), albeit less contaminated by the CIB. Having a noisier SZ map might be a disadvantage if we use auto power spectra or maps, but  this is not an issue if we want to cross-correlate the maps with other tracers (e.g. galaxy densities or weak lensing maps) and we should prioritise maps that are less biased by CIB. For example, the extra SZ power coming from CIB contamination is a potential source of bias for the study of the cosmological parameters from the SZ angular power spectrum \citep{komatsu2002, horowitz2017, douspis2022, tanimura2022a}, and is reduced in our method compared to MILCA, at the cost of slightly poorer precision.

\subsubsection{CMBxSZ}
For the CMBxSZ case (right panel of Fig.~\ref{fig:crossspectra}), no correlation is expected (blue line). We do not find any correlation, either with the recovered components ---indicating that there is no contamination from CMB in the reconstructed SZ map--- or with the SZ in the reconstructed CMB map. This is also the case for the SZ map derived by MILCA; we do not see any correlation with CMB, as expected.

\subsection{SZ fluxes from clusters}

After studying the statistics and contamination of the recovered component maps, we focus on the SZ fluxes from galaxy clusters. For each of the SZ maps, that is, WebSky, MILCA, and $\overline{\mathrm{SZ}}$, we computed the SZ fluxes around a selection of the most massive low-redshift clusters ($\mathrm{M}_{200} > 4.10^{14} \mathrm{M}_\odot$ and $z < 1$)  in
the exact same way ---using aperture photometry--- from the numerical simulation (SZ fluxes for lower masses or higher redshift clusters become too low and noisy). We compare the SZ fluxes  in Fig.~\ref{fig:fluxes}, showing those from MILCA
in green and those from $\overline{\mathrm{SZ}}$  in orange and red. The fluxes displayed in orange are obtained in the training region of the sky while the fluxes displayed in red are obtained in the unseen area of the sky. The very good consistency between all the fluxes indicates a very good recovery of the SZ fluxes for this selection of clusters with all methods, meaning that the network can recover the SZ fluxes well, even in the unseen area, and with an error of the same order of magnitude as that for the fluxes obtained with the traditional state-of-the-art  method MILCA. The dashed blue line in the figure indicates the SZ flux beyond which the clusters have been enhanced in the weight map $w$ used to rebalance the weight of the pixels inside clusters. We see a good correlation even before the vertical line, meaning that even clusters that have not been enhanced are very well recovered. This indicates that the weight map $w$ is helping the network to learn the SZ spectral and spatial features  faster but does not bias the results to the typical objects for which $w$ are input (otherwise, we would see a good agreement only beyond the vertical flux limit and no correlation before). In the right panel of Fig.~\ref{fig:fluxes}, we compare the residuals of the SZ fluxes for MILCA and $\overline{\mathrm{SZ}}$. We see that the fluxes computed in $\overline{\mathrm{SZ}}$ have an error on the same order of magnitude as that for  the fluxes computed with MILCA, with a standard deviation of percentage residuals of $\sigma=40.44\%$ and $\sigma=42.9\%,$ respectively. However, we obtain biased fluxes, which is due to the contamination of CIB. The biases are quite important in MILCA, with 4.97\% bias, while the biases are reduced significantly in the case of $\overline{\mathrm{SZ}}$, with a bias of 1.39\%. These biases obtained with MILCA should be investigated in a more detailed study and confirmed with other standard methods, as they might be an important source of contamination in cosmological analyses using SZ fluxes or SZ spectra.

\section{Discussion and summary}\label{sect:discussion}

We explored a new way of performing component separation using machine learning networks in a self-supervised way. Focusing on the extragalactic submillimetre sky, this method allows us to extract CMB, CIB, and SZ effect maps, and can already be applied to the clean part of the sky (free from foreground emissions), for example to some of the cleanest regions (at very high Galactic latitudes) of the \textit{Planck} frequency maps. Being self-supervised, this method has the potential to be applied directly to data without the need for known labels (the model that is trained here on numerical simulations will not be the one applied to real data). In this paper, we show how we applied our new method to numerical simulations from WebSky and achieved good results when focusing on the power spectra and cross spectra of the different reconstructed components, as well as interesting results regarding the contamination from other components. Our method still has some limitations, which will be investigated in future studies. For example, we did not include a model of the foreground galactic emissions; that is, dust, radio, and CO sources. These emissions could be accounted for by adding other ResUNets  in parallel, allowing the reconstruction of these very same components all at once. We also did not include models of either noise or the effect of the instrumental beams, as the aim of this study is to investigate the theoretical response of these networks in an ideal case and to compare this with the results of a traditional but state-of-the-art method (here, MILCA). Focusing on the individual results for each component: for the CMB case, we obtain a very good reconstruction of the signal, albeit slightly noisy. When we cross-match with the CMB from WebSky, we obtain a very good reconstruction of the signal up to $\ell=2500$ at least. We obtain a very good reconstructed CIB, within a 3\% error and a bias of below 2\% up to $\ell=2500$. For the SZ, we also very nicely recover the signal, with both an advantage and a disadvantage compared to the result obtained with MILCA. Two things can be said about SZ: first, the signal retrieved using our new method is less contaminated by the CIB than that obtain with MILCA. This aspect has very important consequences for the ability to perform cosmology using the SZ signal or to compute cluster masses, and is particularly emphasised in this study. Second, this lower contamination comes with the price of a higher variance, coming from the reconstruction noise that is higher than that obtained with MILCA. Indeed, our method simultaneously constrains CMB, CIB, z, and SZ, and therefore entails a greater number of free parameters than MILCA, which constrains only SZ and removes the CMB. This produces a larger variance in the reconstructed maps at the end. Neither of the reconstructed components, CMB or CIB, is contaminated by other components, as seen in the cross power spectra, and the contamination from CIB in the SZ reconstructed maps is lower than in MILCA, as seen in Fig.~\ref{fig:crossspectra}. To completely remove the contamination of the CIB, our spectral CIB model could be improved to take into account the fact that the emission of the CIB is not a power law but rather the sum of a power law, which could be modelled in greater detail \citep[e.g. in Taylor expansion,][]{chluba2017, vacher2022a, vacher2022b}, but the modelling and reconstruction of the CIB represent a significant challenge, especially for the dust--CIB separation for the detection of the B-modes \citep[e.g.][]{remazeilles2011, remazeilles2018, allys2019, aylor2021, blancard2021}. For example, \cite{allys2019, allys2020} developed an algorithm of wavelet scattering transform \citep[WST,][]{allys2019, blancard2020} similar to convolutional neural networks.The main difference is that the filters of the layers of convolution are fixed instead of learned, and the statistics of the CIB within this transform can later be input as a prior for the extraction of the CIB in component separation.

Compared to other methods using a deep learning network to extract components from CMB data \citep[e.g.][]{caldeira2019, grumitt2020, li2022, wang2022, petroff2020, lin2021}, in our approach we estimate all components simultaneously. This allows a less biased reconstruction, taking into account the dependencies and the correlations with the other components during the training.

The lower effect of the CIB contamination in the power spectra, especially in the reconstruction of the SZ effect, could decrease a potential bias in the computation of the cosmological parameters using the SZ power spectra \citep[][]{salvati2018, douspis2022, gorce2022, tanimura2022a}. Going a step further, and modelling the polarisation, this method could be used in the future for the detection of the B-mode by combining the data of \textit{Planck}, ACT, SPT, SO \citep[][]{ade2019}, and CMB-S4 \citep{cmbs4}.

\begin{acknowledgements} The authors thank the anonymous referee for her/his useful comments that helped increasing the quality of the publication. The authors thank useful discussions with Marc Huertas-Company, and with all the members of the ByoPiC\footnote{\url{https://byopic.eu/team}} project. We also thank Stephane Caminade and IDOC for using the IDOC computing facilities. Part of this research has been supported by the funding for the ByoPiC project from the European Research Council (ERC) under the European Union's Horizon 2020 research and innovation program grant agreement ERC-2015-AdG 695561. 
T.B. acknowledges funding from the French government under management of Agence Nationale de la Recherche as part of the ``Investissements d’avenir'' program, reference ANR-19-P3IA-0001 (PRAIRIE 3IA Institute).
The sky simulations used in this paper were developed by the WebSky Extragalactic CMB Mocks team, with the continuous support of the Canadian Institute for Theoretical Astrophysics (CITA), the Canadian Institute for Advanced Research (CIFAR), and the Natural Sciences and Engineering Council of Canada (NSERC), and were generated on the Niagara supercomputer at the SciNet HPC Consortium (cite https://arxiv.org/abs/1907.13600). SciNet is funded by: the Canada Foundation for Innovation under the auspices of Compute Canada; the Government of Ontario; Ontario Research Fund - Research Excellence; and the University of Toronto.
\end{acknowledgements}

\bibliographystyle{aa}
\bibliography{aanda}

\end{document}